\documentclass[12pt, prd, superscriptaddress,nofootinbib]{revtex4}
\usepackage{amssymb,amsmath,amsfonts}
\usepackage{setspace}
\usepackage{graphicx}
\usepackage{color}

\definecolor{DarkGreen}{rgb}{0,0.4,0}
\definecolor{DeepBlue}{rgb}{0,0,.4}
\usepackage[naturalnames=true,colorlinks=true,citecolor=DarkGreen,urlcolor=blue,linkcolor=DeepBlue]{hyperref}

\begin{document}
\title{Collisional super-Penrose process and Wald inequalities }
\author{I. V. Tanatarov}
\affiliation{Department of Physics and Technology, V.N. Karazin Kharkiv National
University,\\
4 Svobody Square, Kharkiv 61022, Ukraine}
\email{igor.tanatarov@gmail.com}
\author{O. B. Zaslavskii}
\affiliation{Department of Physics and Technology, V.N. Karazin Kharkiv National
University,\\
4 Svobody Square, Kharkiv 61022, Ukraine}
\affiliation{Institute of Mathematics and Mechanics, Kazan Federal University,\\
18 Kremlyovskaya St., Kazan 420008, Russia}
\email{zaslav@ukr.net}

\begin{abstract}
We consider collision of two massive particles in the equatorial plane of an
axially symmetric stationary spacetime that produces two massless particles
afterwards. It is implied that the horizon is absent but there is a naked
singularity or another potential barrier that makes possible the head-on
collision. The relationship between the energy in the center of mass frame $E_{c.m.}$ and the Killing energy $E$ measured at infinity is analyzed. It
follows immediately from the Wald inequalities that unbounded $E$ is
possible for unbounded $E_{c.m.}$ only. This can be realized if the
spacetime is close to the threshold of the horizon formation. Different
types of spacetimes (black holes, naked singularities, wormholes) correspond
to different possible relations between $E_{c.m.}$ and $E$. We develop a
general approach that enables us to describe the collision process in the frames of
the stationary observer and ZAMO (zero angular momentum observer). The
escape cone and escape fraction are derived. A simple explanation of the
existence of the bright spot is given. For the particular case of the Kerr
metric, our results agree with the previous ones found in M. Patil, T. Harada, K.
Nakao, P. S. Joshi, and M. Kimura, Phys. Rev. D \textbf{93}, 104015 (2016). 
\end{abstract}

\maketitle

%\keywords{BSW effect, backreaction force}
%\pacs{04.70.Bw, 97.60.Lf }

\newpage 
\begin{spacing}{1.2}
\tableofcontents
\end{spacing}
\newpage

\section{Introduction}
Several years ago, an interesting observation was made, according to which collision of two particles near the extremal Kerr black hole can lead to formally unbounded energy $E_{c.m.}$ in their center of mass (CM) frame \cite{ban}. Later on, it was shown that this phenomenon has a universal character \cite{prd} and is inherent not only to extremal black holes but also to nonextremal ones \cite{gp}. It is potentially important since it opens new channels of reaction and allows one to probe physics near black holes including processes forbidden in laboratory conditions. A huge series of works followed.

Recently, the emphasis moved from the phenomenon of ultra-high $E_{c.m.}$ to the question of what can be seen by an observer at infinity, who studies physics in a laboratory. In particular, the most important issue is whether the energy at infinity $E$ can be arbitrarily high as well. If yes, then the phenomenon under discussion can contribute to the ultra-high energy cosmic rays observed on Earth \cite{pir1}--\cite{pir3}. The question is, however, nontrivial, since strong redshift can ``eat'' a significant part of the energy excess gain in the collision. Thus high $E_{c.m.}$ does not necessarily imply high energy $E$ at infinity. It turned out that there exist severe restrictions on $E$ that make black holes not prospective in this regard \cite{p}--\cite{omax}. On the other hand, systems without the horizon (in particular, naked singularities) exhibit the possibility of unbounded $E$. For the first time, this was shown in Sec. VII of \cite{head}. Quite recently, a work \cite{ph} appeared in which this phenomenon was investigated in detail for the Kerr overspun metric. The results of \cite{ph} were obtained by means of explicit calculations carried out for this metric.

The problem under discussion is not only important from the theoretical viewpoint. It was pointed out in \cite{Stuchlik1} that superspinning Kerr metric can be relevant for astrophysics and, moreover, exhibit high efficiency of energy extraction in collision processes \cite{Stuchlik2,Stuchlik3}.

In the present paper, we stress that the phenomenon of unbounded $E$ has a deep physical underlying reason and can be simply explained in a couple of lines on the basis of the Wald inequalities \cite{w} extended to the collisional process. Also, we develop general approach in which we derive the relevant features of collisions without specifying the form of the metric. In the particular case of the Kerr metric, our results agree with those of \cite{ph}.

The paper is organized as follows. In section \ref{sec:Wald} we analyze the Wald inequalities and apply them in the context of particle collisions in various scenarios. In Section \ref{sec:frames} we describe the collision in detail in the center of mass frame, the stationary frame, and the LNRF frame and write out the relations between quantities of interest in them. In particular, we derive explicit expressions for the Killing integrals of motion for the massless fragments in terms of $E_{c.m.}$ and the escape angles in the center of mass frame. Then we analyze the limit of large $E_{c.m.}$. The last section \ref{sec:infinity} is devoted to the analysis of particles' geodesic motion, in particular to the possibility of the high-energy particles to escape to infinity in a slightly overspinned naked singularity spacetime. We use the units, in which the fundamental constants are set to unity, $G=c=1$.

One reservation is in order. In our paper we do not consider the effects of the magnetic field, which can, in principle, lead to new interesting consequences. In particular, it was found earlier  \cite{dhu1} that by inclusion of magnetic field around a rotating black hole the Wald and Teukolsky inequalities for the Penrose process are overcome. In \cite{dhu2} it was also pointed out for the first time that the efficiency of the process does not simply increase but can exceed 100\%.  This issue needs further separate investigation.

\section{Energy in CM frame versus Killing energy}
\label{sec:Wald}

In this context two related but different concepts of energy are used. The first one is the energy of the colliding particles in their center of mass frame. It is defined at the event of collision and it determines the collision process, in particular which of the channels of reactions are open and which are not. Second, there are the Killing energies of the colliding particles and of the resulting fragments. The Killing energy of a fragment is its energy as measured at infinity. Therefore, it is of crucial interest whether this quantity can be unbounded. The term ``unbounded'' in this context means that a quantity formally diverges in the limit $N\to 0$, where $N$ is the value of the lapse function at the point of collision. The particular methods to realize this limit vary for different scenarios and depend on the geometry of the spacetime. It is worth noting that the above condition can be easily formulated without a reference to a particular coordinate frame, as $N^2=-\xi^\mu \xi_\mu$ can be expressed covariantly via the two Killing vector fields.

It was obtained in \cite{ph} that the necessary condition for the formally divergent $E$ is that $E_{c.m.}$ should diverge as well (see their Eq. (49) and discussion below). This was obtained for the Kerr metric, for the collision in the equatorial plane. In this Section, we will show that actually this result (in some not quite explicitly articulated form) is already contained in \cite{pir3} (see Eqs. (2.22), (2.57) there) in a quite general setting, and not only for the equatorial motion. We reproduce these results in a more direct way and apply them to the general question regarding the relation between $E$ and $E_{c.m.}$. The issue was not addressed in \cite{pir3} as at
that time the corresponding context was absent. This is discussed below, as we consider different possible relations between potentially divergent quantities, including $E$ and $E_{c.m.}$.

\subsection{Two colliding particles as one compound particle}

As is well known from textbooks, a particle's mass $\mu$ can be found from
the normalization condition for its four-momentum $p^\mu$ as 
\begin{equation}
\mu ^{2}=-p_{\mu }p^{\mu };  \label{mu}
\end{equation}
for massless particles $p^\mu$ should be replaced with the wavevector $%
k^{\mu }$.

Let us consider the collision of a pair of particles with momenta $p_{1}^\mu$
and $p_2^\mu$. By analogy with the above, we can define the energy in the
center of mass frame of the pair in terms of its total four-momentum at the
collision event $P^{\mu }=p_{1}^{\mu }+p_{2}^{\mu}$, as 
\begin{equation}
M^{2}=-P_{\mu }P^{\mu }.
\end{equation}
Using this notation one can always imagine the instantaneous system composed
of one compound particle of mass $M\equiv E_{c.m.}$ and momentum $P^{\mu}$
as an intermediate state in any collision process. Although this depiction
is only valid at the collision event, this is quite sufficient to derive
some important general relations.

%\subsection{From Wald inequalities to collisional Penrose process}

\subsection{General features of collisional Penrose process from Wald
inequalities}

In this section we show that some of the crucial properties of the
collisional Penrose process follow quite straightforwardly from Wald's
inequalities.

\subsubsection{Two massless fragments}

The Wald inequality \cite{w} gives the range of possible energies for a
fragment in the Penrose process resulting from the decay of a particle in
the ergosphere of a rotating black hole. For the case when a particle with
mass $\mu $ and Killing energy $E$ decays into two massless fragments, the
inequality is for the ratio of the fragment's frequency measured at infinity 
$\omega _{\infty }$ to its emitted frequency $\omega $ measured in the rest
frame of the decaying particle (see formula in the footnote of \cite{w}): 
\begin{equation}
\frac{E}{\mu }-\sqrt{\frac{E^{2}}{\mu ^{2}}+g_{tt}}\leq \frac{\omega
_{\infty }}{\omega }\leq \frac{E}{\mu }+\sqrt{\frac{E^{2}}{\mu ^{2}}+g_{tt}}.
\label{WGamma}
\end{equation}
Here and below $g_{tt}$ is the value taken at the decay event.

For the collisional Penrose process we substitute $E=E_{1}+E_{2}$ and $\mu
=M $, which for the two photons with frequency $\omega $ emitted in opposite
directions in the CM frame is $M=2\hbar \omega $, so (\ref{WGamma}) implies 
\begin{equation}
E-\sqrt{E^{2}+g_{tt}M^{2}} \leq 2\hbar \omega _{\infty } \leq E+\sqrt{%
E^{2}+g_{tt}M^{2}}.  \label{W2Gamma}
\end{equation}%
%
%
%
%As explained above, (\ref{WaldMassless}) can be applied straightforwardly for the collisional Penrose process by substituting $\mu =M$ and $E=E_{1}+E_{2}$.

We see that for finite $g_{tt}$ and $E$, the frequency measured at infinity $%
\omega _{\infty }$ is unbounded only if $E_{c.m.}\equiv M$ also diverges.
The situation when both quantities are unbounded is possible in the
background of spacetimes with naked singularity and no horizon \cite{head}, 
\cite{ph}. An example of the case when $M$ diverge, while $\omega_\infty$
stays finite, is the collisional Penrose process near black holes (see
discussion in Introduction above). Thus the divergence of $M$ is necessary,
but not sufficient for the divergence of $\omega _{\infty }$.

In the limit $M\rightarrow \infty $ the maximum possible value of the
fragment's energy at infinity is 
\begin{equation}  \label{max}
\left(E_\infty \right)_{max} \equiv \hbar \left( \omega _{\infty }\right)
_{\max }\approx \sqrt{g_{tt}}\;\frac{M}{2}.
\end{equation}

\subsubsection{Two massive fragments}
The process of a particle's decay into two massive fragments in its center
of mass frame is described by three independent parameters, such as the two
fragments' masses and their total energy, while the escape angles only
affect how this process is seen by other observers and, by extension, the
fragments' Killing energies. If one of the fragments with mass $M^{\prime }$
has velocity $v$ and Lorentz factor $\gamma $ in the frame of the decaying
particle, then, following Sec. 7.65 of \cite{ch}, one can derive the
expression for the Killing energy $E^{\prime}$. For a reader's convenience,
we outline briefly the derivation below.

One can express the four-velocity of a decaying particle $u$ and the time-like
Killing vector $\partial_t =\xi$ in terms of an orthonormal tetrad $e_{(i)}^{\mu}$ and $U^{\mu}$, where $U^\mu$ is the timelike unit vector, as
\begin{align}
	u^{\mu }&=\gamma \lbrack U^{\mu }+v^{(i)}e_{(i)}^{\mu }],\\
	\xi_{\mu }&=\xi _{(0)}U_{\mu }+\xi _{(i)}e_{\mu }^{(i)}.
\end{align}
In terms of these tetrad components the quantities of interest are
\begin{align}
	g_{tt}=&\xi _{\mu }\xi ^{\mu }=-\xi _{(0)}^{2}+|\xi |^{2};\\
	\frac{E}{\mu }=&-\xi _{\mu }U^{\mu }=\xi _{(0)};\\
	\frac{E^{\prime }}{M^{\prime }}
		=&-u_{\mu }\xi ^{\mu }=\gamma \big[\xi _{(0)}-\xi _{(k)}v^{(k)}\big];\label{E'/M'}\\
	&\xi _{(k)}v^{(k)}=-|\xi|\, |v|\;\cos \theta ,
\end{align}
where $|\xi|$ and $|v|$ are the lengths of spatial vectors $v^{(i)}$ and $\xi^{(i)}$. The minus sign in the definition of angle $\theta$ between the spatial components of $u$ and $\xi$ is chosen for convenience. On substituting everything into Eq. (\ref{E'/M'}), we get
\begin{equation}
	\frac{E^{\prime }}{M^{\prime }}
		=\gamma \frac{E}{\mu }
			+\gamma |v|\;\sqrt{\frac{E^{2}}{\mu ^{2}}+g_{tt}}\;\cos \theta   ,\label{ChM}
\end{equation}
which corresponds to Eq. (328) of Sec. 7.65 of \cite{ch}.

In our case the decaying particle is replaced with the effective compound
one as explained above, so $v$, $\gamma $, and $\theta $ are measured in the
center of mass frame of the colliding particles at the collision.

For two identical fragments with masses $m$ we have $\mu =M=2\gamma m$, so,
in terms of a fragment's Lorentz factor at infinity $\gamma _{\infty} =E^{\prime}/M^{\prime}$, Eq. (\ref{ChM}) implies 
\begin{equation}
E-|v|\sqrt{E^{2}+g_{tt}M^{2}}\leq 2m\gamma _{\infty } \leq E+|v|\sqrt{%
E^{2}+g_{tt}M^{2}}.  \label{ChMRange2}
\end{equation}
Once again the observed energy at infinity can be unbounded only if the energy in
the center of mass frame $M$ diverges. In this limit, $M\rightarrow \infty $, the maximal energy at infinity is 
\begin{equation}  \label{ginf}
\left(E_\infty \right)_{max}\equiv m\left(\gamma _{\infty
}\right)_{max}\approx |v|\sqrt{g_{tt}}\;\frac{M}{2}.
\end{equation}

In the particular case of the Kerr metric, this gives us the results of \cite{ph}, obtained there by direct calculations with transformations between different frames. In particular, the general statement\textemdash that for $\gamma_\infty$ to be unbounded $M$ also has to diverge\textemdash agrees with Eq. (49) and subsequent discussion of \cite{ph}.

\subsection{Diverging metric coefficient}

In deriving the above conclusions, we tacitly assumed that all metric coefficients remain finite. There is a case of interest, however, when $g_{tt}\rightarrow \infty $. This corresponds to rapidly rotating spacetimes, with formally divergent $\omega \rightarrow \infty $ (see Eqs. (\ref{Metric}), (\ref{gtt}) for the metric below). Astrophysical significance of this situation is still unclear, but it can happen, in particular, for wormholes \cite{wh1}--\cite{rotgrg}. In this limit both for massless (\ref{W2Gamma}) and massive (\ref{ChMRange2}) fragments we find that the maximum energy at infinity is given by the same formulae (\ref{max}) and (\ref{ginf}), so that 
\begin{equation}
	\left(E_{\infty}\right)_{max}\sim \sqrt{g_{tt}}\to \infty
\end{equation}
even when $M$ is finite. We see that this quite unusual result, obtained in the aforementioned papers by detailed analysis of the conservation laws, can be simply explained with the help of the Wald inequalities.

One reservation is in order here. When $\omega$ becomes larger and larger, the components of the curvature tensor increase and the geometry approaches the singular case. However, one can choose an intermediate range of parameters in which both the energy in the center of mass and the curvature are very large but still in the classical domain, much less than the corresponding Planck scales. See Sec. 4 of \cite{rotmpla} for details.

\subsection{BSW effect with finite energy at infinity}

For collisional Penrose process near black holes it was shown, that even when the BSW effect takes place (so that $M$ diverge), the efficiency of the Penrose process remains finite, as $E_\infty$ remains bounded \cite{pir3,p,j,z}. This case corresponds to $\mu=M\rightarrow \infty $, $\gamma \rightarrow \infty $, $v\rightarrow 1$ in Eq. (\ref{ChM}), so we have 
\begin{equation}
	E_\infty \approx \sqrt{g_{tt}}\;\frac{M}{2}\cos \theta .
\end{equation}
For the left hand side to be bounded, the escape velocities for those particles that reach infinity must be restricted to an (infinitely) narrow cone, with $\cos \theta \rightarrow 0$.

\section{Collision in LNRF and CM frames}
\label{sec:frames}

\subsection{A particle in rotating BH spacetime}

We consider the motion of particles in the equatorial plane of an axially
symmetric stationary spacetime. Its metric can always be brought to the form 
\begin{equation}  \label{Metric}
ds^2 =-N^2 dt^2 +g_\phi (d\phi -\omega dt)^2 +\alpha^2 \frac{dr^2}{N^2}.
\end{equation}
There are two Killing vectors, $\partial_t$ and $\partial_\phi$, so all
metric functions depend only on the radial coordinate $r$. The factor $%
\alpha $ can be chosen to be equal to $1$ everywhere in the equatorial plane
by redefining the radial coordinate. We will also use notation 
\begin{equation}  \label{gtt}
g_t \equiv g_{tt}=g_\phi \omega^2 -N^2
\end{equation}
for the $tt$ component of the metric tensor. Although the considered
spacetime is effectively three-dimensional, we will use the usual terms
4-velocity, tetrad, etc.

A particle's 4-velocity can be parametrized by its integrals of motion, $E$
and $L$, in the following way: 
\begin{equation}  \label{Part4velocity}
u_\mu =\Big(\!-E,L,\frac{\alpha^2}{N^2}Z\Big),\qquad u^\mu =\Big(\frac{X}{N^2%
},\frac{\omega X}{N^2}+\frac{L}{g_\phi},Z\Big).
\end{equation}
Here 
\begin{equation*}
X=E-\omega L,
\end{equation*}
while $Z$ (which can be both positive and negative) is determined from the
normalization condition: 
\begin{equation}  \label{Z}
\alpha^2 Z^2 =X^2 -N^2 \Big(\frac{L^2}{g_\phi}+\epsilon^2\Big).
\end{equation}
For massive particles $\epsilon^2 =1$, while for massless ones $\epsilon^2
=0 $. Hereafter we will assume $\alpha=1$.

For massive particles $E$ and $L$ are the energy and angular momentum per
unit mass, and for massless particles the worldline parametrization can be
chosen so that they are either energy and angular momentum, or unity and
impact parameter respectively.

\subsection{Particles in locally non-rotating frame}

%% LNRF/ OZAMO frame

Let us consider the collision of two particles with 4-velocities $%
u^\mu_{1,2} $, integrals of motion $E_{1,2}$ and $L_{1,2}$, and respective
values of parameters $X_{1,2}$ and $Z_{1,2}$ at the event of collision. For
informative description of the process it is convenient to use the locally
non-rotating frame (LNRF), attached to observers with zero angular momenta,
orbiting at constant $r$ (also called orbiting zero angular momentum observers, OZAMOs) 
\cite{72}. The 1-form tetrad of this frame is 
\begin{equation}  \label{LNRFtetrad}
e_o^{(t)}=N\,dt,\qquad e_o^{(\phi)}=\sqrt{g_\phi}(d\phi -\omega dt),\qquad
e_0^{(r)}=\frac{dr}{N}.
\end{equation}
The tetrad components of the two particles' 4-velocities (\ref{Part4velocity}) are then 
\begin{equation}  \label{UTetrad}
	u^{(i)}_a=u^\mu_a (e_o^{(i)})_\mu 
		= \Big(\frac{X_a}{N},\frac{L_a}{\sqrt{g_\phi}},
			\frac{Z_a}{N}\Big),\qquad a=1,2.
\end{equation}
and the total 4-momentum of the pair at the collision event is 
\begin{equation}
P^{(i)}\equiv (\mathcal{E},P_\phi ,P_r) =m_1 u_1^{(i)}+m_2 u_2^{(i)}.
\end{equation}

Its components are 
\begin{align}
\mathcal{E}&\equiv \frac{\mathrm{X}_0}{N};  \label{A} \\
P_\phi&\equiv \frac{\mathrm{L}_0}{\sqrt{g_\phi}} ;  \label{C} \\
P_r&\equiv \frac{\mathrm{Z}_0}{N},  \label{B}
\end{align}
where (for a massless particle we should set $m_i =1$ and replace $u^\mu$
with $k^\mu$) 
\begin{align}
\mathrm{L}_{0}&=m_1 L_{1}+m_2 L_{2};  \label{L0} \\
\mathrm{E}_{0}&=m_1 E_{1}+m_2 E_{2};  \label{E0} \\
\mathrm{X}_{0}&=m_1 X_{1}+m_2 X_{2}=\mathrm{E}_0 -\omega \mathrm{L}_0;  \label{X0} \\
\mathrm{Z}_0&=m_1 Z_1 +m_2 Z_2 .  \label{Z0}
\end{align}
Here $L_{1,2}$ and $E_{1,2}$ are the Killing energies of the particles per
unit mass, and $\mathrm{E}_0$ and $\mathrm{L}_0$ (in Roman script) are the total additive integrals of motion of the pair, while $\mathrm{X}_0$ and $\mathrm{Z}_0$ are only defined in the collision event; all values of metric functions are taken at the collision event.

The total three-momentum $P$ and the mass, associated with energy $\mathcal{E}$ and momentum $P$, are 
\begin{align}
P^2&=P_r^2 +P_\phi^2;  \label{P} \\
M^2 &=\mathcal{E}^2 -P^2 .  \label{M}
\end{align}
In case the result of collision was the production of a single particle, its
mass would be $M$. In the general case it is the energy in the center of
mass frame of the fragments. It is clear from Eqs. (\ref{B}), (\ref{Z0}) and (\ref{M}) that for the same energies and angular momenta of the colliding
particles, the energy in the center of mass frame is greater when they are
traveling in the opposite directions in the radial coordinate. Indeed, when $Z_1 Z_2
<0$, the absolute values of $Z_0$ and $P$ are less than when $Z_1 Z_2 >0$, while $\mathrm{X}_0$ and $\mathcal{E}$ stay the same, which leads to greater $M$ in (\ref{M}). This is a general property, not restricted to the case of the Kerr metric \cite{ph}.

If the collision happens in the region where $N$ is small, $N\ll 1 $, the formal
limit $N\to 0$ yields the following asymptotic behavior: 
\begin{align}
\mathcal{E}&\approx\frac{\mathrm{E}_0 -\omega \mathrm{L}_0}{N}; \\
P_r &\approx \frac{m_1 X_1 \sigma_1 + m_2 X_2 \sigma_2 }{N},
\end{align}
where 
\begin{equation}
	\label{sigma}
	\sigma_a =\text{sign}Z_a .
\end{equation}
All the metric functions other than $N$ are taken here at the point of
expansion: if there is a horizon, where $N$ turns to zero, this is the
horizon, for a slightly overspinned naked singularity spacetime this will be
the point where $N^2$ reaches the minimum.

Suppose the two colliding particles have the same mass $m$ and fall from
infinity with zero velocity, so that $E_1 = E_2 =1$ and $\mathrm{E}_0 =2m$, then one
of them is reflected from the effective potential below the small $N$ region
and while on the outgoing trajectory collides with the ingoing second
particle. This is the regime in which $M$ is maximized, as noted above. In
this case $\sigma_1=+1$, $\sigma_2 =-1$, and 
\begin{align}
\mathcal{E}&\approx m\frac{2-\omega (L_1 +L_2)}{N};  \label{Eappr} \\
P_{r}&\approx m\frac{X_{1}-X_{2}}{N}=m\frac{\omega(L_{2}-L_{1})}{N}.
\end{align}
This is the generalization of Eq. (70) of \cite{ph} ($P_\phi$ is still given
by Eq. (\ref{C})).

As discussed above, one can imagine any collision process going through an
intermediate virtual stage, in which there is only one particle present,
carrying the total energy and momentum of the system. Therefore the two
quantities, $M$ and $\mathcal{E}$, completely determine the initial
conditions for the scattering event in the center of mass frame, except for the escape angles, which depend on the details of the interaction.

We will see that the unbounded growth of the energy of collision in the center of mass frame is due to the geometry of spacetime (the horizon is not formed but is on the threshold of its formation), so our assumption about equal masses of colliding particles is not restrictive and does not affect the overall picture.

\subsection{From LNRF to center of mass frame}

The collision process looks simplest in the center of mass (CM) frame. This
frame's movement with respect to the LNRF frame is characterized by velocity 
$V =P/\mathcal{E}$, or equavalently rapidity $V=\tanh\chi$, and the angle $%
\psi$ between this velocity and the radial direction: 
\begin{align}
	&\cosh\chi =\frac{\mathcal{E}}{M},\quad \sinh\chi=\frac{P}{M};
		  \label{chi}\\
	&\cos\psi=\frac{P_r}{P},\quad \sin\psi=\frac{P_\phi}{P}.  
		\label{psi}
\end{align}
Then the transformation matrix from the LNRF frame to the CM frame for the
tetrad components of a 4-vector $A^{(i)}$ is the product of rotation and
boost: 
\begin{equation}  \label{Lambdas}
A_{cm}=\Lambda_{o\to cm} A_{o},\qquad \Lambda_{o\to cm}
=\Lambda_{boost}\Lambda_{rot},
\end{equation}
where the boost and rotation matrices are given by 
\begin{align}  \label{Matrices}
\Lambda_{boost}=%
	\begin{pmatrix}
		\cosh \chi & 0 & -\sinh\chi \\ 
		0 & 1 & 0 \\ 
		-\sinh\chi & 0 & \cosh\chi%
	\end{pmatrix};
\qquad
&\Lambda_{rot}=
	\begin{pmatrix}
		1 & 0 & 0 \\ 
		0 & \cos\psi & -\sin\psi \\ 
		0 & \sin\psi & \cos\psi%
	\end{pmatrix}.%
\end{align}

The explicit direct and reverse transformation matrices then take form 
\begin{align}
\label{MatrixO-CM}
\Lambda_{o\to cm} &=%
\begin{pmatrix}
\cosh \chi & -\sinh\chi \sin\psi & -\sinh \chi \cos\psi \\ 
0 & \cos\psi & -\sin\psi \\ 
-\sinh\chi & \cosh\chi \sin\psi & \cosh\chi \cos\psi%
\end{pmatrix}%
; \\
\label{MatrixCM-O}
\Lambda_{cm\to o}\equiv \Lambda^{-1}_{o\to cm} &=%
\begin{pmatrix}
\cosh\chi & 0 & \sinh\chi \\ 
\sinh\chi \sin\psi & \cos\psi & \cosh\chi \sin\psi \\ 
\sinh\chi \cos\psi & -\sin\psi & \cosh\chi \cos\psi%
\end{pmatrix}.
\end{align}

\subsection{Collision in CM frame: reaction products}
\label{sec:LE3}

If the collision produces two massless particles, their 4-momenta in the CM
frame will be 
\begin{align}
p_{3\,cm}^{(i)}&=\frac{M}{2}(1,+\sin\theta,+\cos\theta); \\
p_{4\,cm}^{(i)}&=\frac{M}{2}(1,-\sin\theta,-\cos\theta).
\end{align}
Here $\theta$ is the angle between their momenta and the radial direction,
it is the free parameter of the collision process. Hereafter we look at the first of the two new
particles, with the plus sign and subscript ``3'', consider its energy and
the possibility of its escape to infinity.

The particle's momentum in the LNRF frame is (see (\ref{UTetrad}) and
definitions (\ref{L0})--(\ref{Z0})) 
\begin{equation}
p^{(i)}_{3\,o}=\Lambda^{-1}_{o\to cm}\,p_{3\,cm}^{(i)}
=\Lambda_{rot}^{-1}\Lambda_{boost}^{-1}\; p_{3\,cm}^{(i)} =\Big(\frac{X_3}{N}%
,\frac{L_3}{\sqrt{g_\phi}},\frac{Z_3}{N}\Big),
\end{equation}
with matrices $\Lambda_{rot}$ and $\Lambda_{boost}$ from (\ref{Matrices}),
from which we obtain for its components 
\begin{align}
\frac{X_3}{N}&=\frac{M}{2}\big[\cosh\chi +\sinh\chi \cos\theta \big]; 
	\label{X30}\\
\frac{L_3}{\sqrt{g_\phi}}&=\frac{M}{2} \big[\sinh\chi \sin\psi 
	+\cosh\chi \sin\psi \cos\theta +\cos\psi \sin\theta \big].
	\label{L30}
\end{align}

Using Eqs. (\ref{chi})--(\ref{psi}) and expressing the boost and rotation
angles through the parameters of the collision (\ref{A})--(\ref{B}), we
obtain 
\begin{equation}  \label{X3}
	X_3=\frac{\mathrm{X}_0 }{2}+N\frac{P}{2}\cos\theta
\end{equation}
and for the angular momentum 
\begin{align}
	L_3 =&\frac{\mathrm{L}_0}{2}
		+\frac{\Lambda}{2}\sin (\theta-\theta_L);  \label{L3} \\
	&\Lambda =\sqrt{\mathrm{L}_0^2 +g_\phi M^2}; 
			\label{L3ampl} \\
	&\sin\theta_L =-\frac{\mathcal{E}}{P}\cdot \frac{\mathrm{L}_0}{\Lambda};
			\label{SinTL} \\
	&\cos\theta_L =\frac{P_r}{P}\cdot \frac{\sqrt{g_\phi}\;M}{\Lambda}.
			\label{CosTL}
\end{align}

Then the energy of the particle is $E_3=X_3 +\omega L_3$, and after some
lengthy but elementary algebra the expression can be brought to the
following simple form: 
\begin{align}
	E_3 &=\frac{\mathrm{E}_0}{2}+\frac{W}{2}\sin (\theta-\theta_E);  \label{E3} \\
	&W=\sqrt{\mathrm{E}_0^2 +g_t M^2};  \label{E3ampl} \\
	&\sin\theta_E =-\frac{\mathrm{E}_0 \mathrm{X}_0-N^2M^2}{NPW};  
			\label{SinTE} \\
	&\cos\theta_E =\frac{P_r}{P}\cdot \frac{\omega \sqrt{g_\phi}\;M}{W}.
			\label{CosTE}
\end{align}
This is the much more general and simple form of the formulas (40) and (45)
of \cite{ph} for the Kerr metric (with our $\theta$ denoted in \cite{ph} as $%
\alpha$).

We see that the expression for $E_3$ is the analogue of Eq. (\ref{ChM}),
written for massive fragments, and it implies the Wald's inequality (\ref%
{W2Gamma}) for massless ones.

Below we will also make use of the angle difference 
\begin{align}  \label{ThetaPrime}
\theta^{\prime}=&\;\theta_E-\theta_L ; \\
&\sin\theta^{\prime}=-\frac{\sqrt{g_\phi}\; \mathrm{Z}_0 M}{W\Lambda};  
	\label{SinT'}\\
&\cos\theta^{\prime}
	=\frac{\mathrm{L}_0 \mathrm{E}_0 +\omega g_\phi M^2}{W\Lambda}.
	\label{CosT'}
\end{align}

\subsection{Large $M$ limit}
\label{sec:limit}

Suppose $E_3 \to \infty$. Then $W\to \infty$ and $M\to \infty$. The observed
energy of the created particle can only be divergent if the energy in the
center of mass is. Let us consider the asymptotic behavior of different
quantities in the limit 
\begin{equation}  \label{Mlimit}
	M^2 =\frac{\mathrm{X}_0^2 -\mathrm{Z}_0^2 }{N^2}-P_\phi^2 \to +\infty,
\end{equation}
where we have used (\ref{A})--(\ref{B}) and (\ref{M}). If we consider
generic particles, with no fine-tuning, then $\mathrm{X}_0$ and $\mathrm{Z}_0$ are both separated from zero and have different limits, so\footnote{If one of the particles is fine-tuned (the so-called critical particle, see e.g. \cite{prd} for more details), which is necessary for the BSW effect, then $\mathrm{X}_0^2 -\mathrm{Z}_0^2 \sim N$ and the asymptote is different $M\sim N^{-1/2}$. We will not consider this case here.} 
\begin{equation}
	M\sim \frac{1}{N}\to +\infty.
\end{equation}

\paragraph{Amplitudes.}

For the amplitudes $\Lambda$ (\ref{L3ampl}) and $W$ (\ref{E3ampl}) we have 
\begin{align}
	\Lambda& =\sqrt{g_\phi}\; M 
		\Big(1+\frac{\mathrm{L}_0^2}{g_\phi}\,\frac{1}{2M^2}+O(N^4)\Big); 
			 \label{LambdaN} \\
	W&= \omega \sqrt{g_\phi}\; M 
		\Big(1 +\frac{\mathrm{E}_0^2 +\mathrm{Z}_0^2 -\mathrm{X}_0^2}
			{g_\phi \omega^2}\,\frac{1}{2M^2} +O(N^4)\Big).  \label{WN}
\end{align}

From (\ref{X3}) we see that $X_3$ remains finite: $\mathrm{X}_0$ is bounded, and $P\sim P_r \sim 1/N$, so in the zeroth order by $N$ it is equal to 
\begin{equation}  
	\label{X3appr}
	X_3 \approx \frac{\mathrm{X}_0}{2}+\frac{|\mathrm{Z}_0|}{2}\cos\theta.
\end{equation}

\paragraph{Angles.}

Of the trigonometric functions of angles $\theta_L$ and $\theta_E$ it is
sufficient to derive only two asymptotes, while bearing in mind all the
signs: from (\ref{SinTL}) and (\ref{CosTL}) we have 
\begin{align}
	\sin\theta_L &
		=-\frac{\mathrm{X}_0}{M} 
			\frac{P_\phi}{\sqrt{\mathrm{Z}_0^2 +P_\phi^2 N^2}} 
			\Big(1-\frac{\mathrm{L}_0^2}{g_\phi}\,\frac{1}{2M^2}+O(N^4) \Big) \sim N ; \\
	\cos\theta_L &\approx \text{sign}(\mathrm{Z}_0).
\end{align}
Likewise, using also (\ref{SinT'}) and (\ref{CosT'}), we derive the expansions for $\theta^{\prime}$ in the following form: 
\begin{align}
	\sin\theta^{\prime}&
		= -\frac{\mathrm{Z}_0}{M}\;\frac{1}{\omega \sqrt{g_\phi}} 
			\Big(1-\frac{\mathrm{Z}_0^2
				+2\omega \mathrm{E}_0 \mathrm{L}_0}{g_\phi \omega^2}
					\, \frac{1}{2M^2} +O(N^4) \Big)   \\
	&= -\frac{N}{\omega\sqrt{g_\phi}} \;v_0 \gamma_0 \cos\psi \sim N \\
	 \cos\theta^{\prime} 
		& \approx +1,\qquad \theta_E \approx \theta_L .
\end{align}

Thus all angles $\theta_L, \theta_E, \theta^{\prime}$ are small and of the
order of $N$. There is a narrow cone with the angle $O(N)$ around the radial
direction $\theta=0$, in which $E_3$ and $L_3$ are finite despite diverging $%
M$, while for all other angles both $E_3$ and $L_3$ diverge as $M\sim 1/N$.

In the generic situation $\mathrm{Z}_0 \neq 0$, so 
\begin{equation}
	\label{SmallAngles}
	\theta_L \sim \theta^{\prime}\sim \theta_E \sim N ;
\end{equation}

There is also a special case, corresponding to such fine-tuning of the
particles' parameters, that the composite particle has zero radial velocity, 
$\mathrm{Z}_0 =0$ and $P_r =0$. In this case from (\ref{SinT'}) and (\ref{CosTL}) we
find immediately that $\theta^{\prime}=0$ and $\cos\theta_L =0$ exactly, and
taking into account the signs, we have 
\begin{equation}
\sin \theta_L =\sin \theta_E =-1 ,
\end{equation}
exactly.

\paragraph{Impact parameter.}

When both $\Lambda$ and $W$ diverge, and $\theta$ is not in the narrow cone $%
\sim O(N)$, the impact parameter tends to 
\begin{equation}
b_3 \equiv \frac{L_3}{E_3}\approx \frac{\Lambda}{W}\approx \frac{1}{\omega},
\end{equation}
as follows from (\ref{L3}), (\ref{E3}), (\ref{LambdaN}), and (\ref{WN}). In
order to find the asymptotic behavior in the next order of magnitude, we use
the expansions for all quantities both in numerator and denumerator,
retaining terms of the order of $N^2$, and after some algebraic
transformations obtain 
\begin{align}
b_3 &=\frac{1}{\omega}\Big[1-\frac{N}{\omega \sqrt{g_\phi}} \frac{\mathcal{E}%
+P_r \cos(\theta-\theta_E)}{M\sin (\theta-\theta_E)}+O(N^2)\Big]
\end{align}
All the metric coefficients here are taken at the point of collision.

It appears to be more convenient below to use the inverse impact parameter 
\begin{equation}
\beta_3 \equiv \frac{E_3}{L_3}= b_3^{-1} =\omega +\frac{N}{\sqrt{g_\phi}} 
\frac{\mathcal{E}+P_r \cos(\theta-\theta_E)}{M\sin (\theta-\theta_E)}+O(N^2).
\end{equation}
As in the considered limit $P_r =P (1+O(N^2))$, and $\theta_E \sim N$, when
the angle is not small, $\theta_E$ can be omitted, and this expression can
be brought to 
\begin{align}
\beta_3 &=\omega +\frac{N}{\sqrt{g_\phi}} \;\frac{\mathcal{E}+P\cos\theta}{%
M\sin\theta}+O(N^2) .
\end{align}

It is worth noting that both the energy and spin of the central gravitating compact object (black hole, singularity or overspinned black hole) changes as a result of the processes we considered. However, as long as its overall structure is preserved, e.g. a naked singularity does not convert into a black hole, the results should remain valid. More detailed treatment would include backreaction of particles on the metric itself but this very interesting (although quite difficult) task is beyond the scope of our paper and can be a subject of future research.

\subsection{Parameters of colliding particles}
Let us consider the 4-velocity of one of the colliding particles in the center of mass frame 
\[u_{a\,cm}^{(i)}=(u_{a\,cm}^{(t)},u_{a\,cm}^{(\phi)}, u_{a\,cm}^{(r)}), \qquad a=1,2,\]
in terms of its integrals of motion. The particle's 4-velocity is given by (\ref{Part4velocity}) and its tetrad components in the LNRF frame (\ref{UTetrad}).  Using the transformation matrix (\ref{MatrixO-CM}) between the LNRF and CM frames, for its 4-momentum in the CM frame we get
\begin{align}
	u_{a\,cm}^{(t)} &= \cosh \chi \;\frac{X_a}{N}
		-\sinh\chi \sin\psi \;\frac{L_a}{\sqrt{g_\phi}}
			-\sinh\chi \cos\psi \;\frac{Z_a}{N};\\
	u_{a\,cm}^{(\phi)}&=\cos\psi \;\frac{L_a}{\sqrt{g_\phi}}-\sin\psi \;\frac{Z_a}{N};\\
	u_{a\,cm}^{(r)}&=-\sinh\chi \;\frac{X_a}{N}
		+\cosh\chi \sin\psi \;\frac{L_a}{\sqrt{g_\phi}}
			+\cosh\chi \cos\psi \;\frac{Z_a}{N}.
\end{align}
These formulae are the analogues of Eqs. (72)--(74) of \cite{ph}.

Let there be no fine-tuning, so that $\mathrm{Z}_0$ (\ref{Z0}) is not small and $P_r \sim N^{-1}$ (\ref{B}) is large. Then $\mathcal{E},M,P,P_r\sim N^{-1}$ are all large, while $L_a = O(1)$, so $\psi$ (\ref{psi}) is close to either zero or $\pi$, depending on the sign of $P_r$, which determines the direction of motion of the composite particle in the radial direction: 
\[\cos\psi\approx \text{sign}(P_r) 
	=\text{sign}(\mathrm{Z}_0) \equiv\sigma =\pm 1.\]
In this limit
\begin{align}
	u_{a\,cm}^{(t)} &\approx \frac{X_a \cosh \chi -\sigma Z_a \sinh\chi}{N};\\
	u_{a\,cm}^{(\phi)} & 
		= \sigma\frac{L_a}{\sqrt{g_\phi}}-\frac{P_\phi}{P}\frac{Z_a}{N};\\
	u_{a\,cm}^{(r)}&\approx \frac{\sigma Z_a \cosh\chi - Z_a \sinh\chi}{N}.
\end{align}
For usual particles, with $X_a$ not fine-tuned to be small, $Z_a =\sigma_a X_a +O(N^2)$, where $\sigma_a=\text{sign} Z_a$ (\ref{sigma}), therefore
\begin{align}
	u_{a\,cm}^{(t)}  &\approx \frac{X_a}{N}[\cosh\chi -\sigma \sigma_a \sinh\chi];\\
	u_{a\,cm}^{(r)}&\approx \sigma \sigma_a u_{a\,cm}^{(t)} . \label{ur}
\end{align}
It can be verified that $m_1 u_{1\,cm}^{(t)}$ and $m_2 u_{2\,cm}^{(t)}$ add up to $M$ in this limit, as they should.

Taking into account (\ref{A})--(\ref{Z0}), for the $\phi$ component of particle 1's velocity  we have
\begin{equation}
	u_{1\,cm}^{(\phi)} =g_\phi^{-1/2}\frac{L_1 \cdot m_2Z_2 - L_2 \cdot m_1 Z_1}
		{|m_1 Z_1 +m_2 Z_2|}.
\end{equation}

\section{High-energy particles at infinity}
\label{sec:infinity}

\subsection{The whale diagram and escape cones}

It is often convenient to analyze a particle's motion in an axially
symmetric stationary spacetime in terms of angular velocities. Let a
particle's angular velocity be 
\begin{equation}
\Omega=\frac{d\phi}{dt}=\frac{u^\phi}{u^t}.
\end{equation}
Then writing the normalizing condition for the four-velocity in the form ($%
\epsilon^2=0$ for massless particles and $\epsilon^2=1$ for massive ones) 
\begin{equation}
-\epsilon^2=(u^t)^2 \big[g_t +2\Omega g_{t\phi}+\Omega^2 g_{\phi}+g_{rr}
v_r^2 \big],
\end{equation}
where $v_r =dr/dt$, we obtain the possible range for the angular velocity $%
\Omega$ from the inequality 
\begin{equation}
(\Omega-\omega)^2 \leq \frac{N^2}{g_\phi},
\end{equation}
which becomes equality only for massless particles with zero radial
velocity. Then 
\begin{equation}
\Omega \in (\Omega_- ,\Omega_+), \qquad \Omega_{\pm}=\omega \pm
\delta\Omega, \qquad \delta \Omega =\sqrt{\frac{N^2}{g_\phi}},
\end{equation}
The three metric functions $\omega$ and $\Omega_\pm$ can be called ``natural
angular velocities'', as they define a particle's kinematics in the rotating
stationary spacetime.

\begin{figure}[htb]
\center
\includegraphics[width=0.49\textwidth,viewport=90 487 519 758]{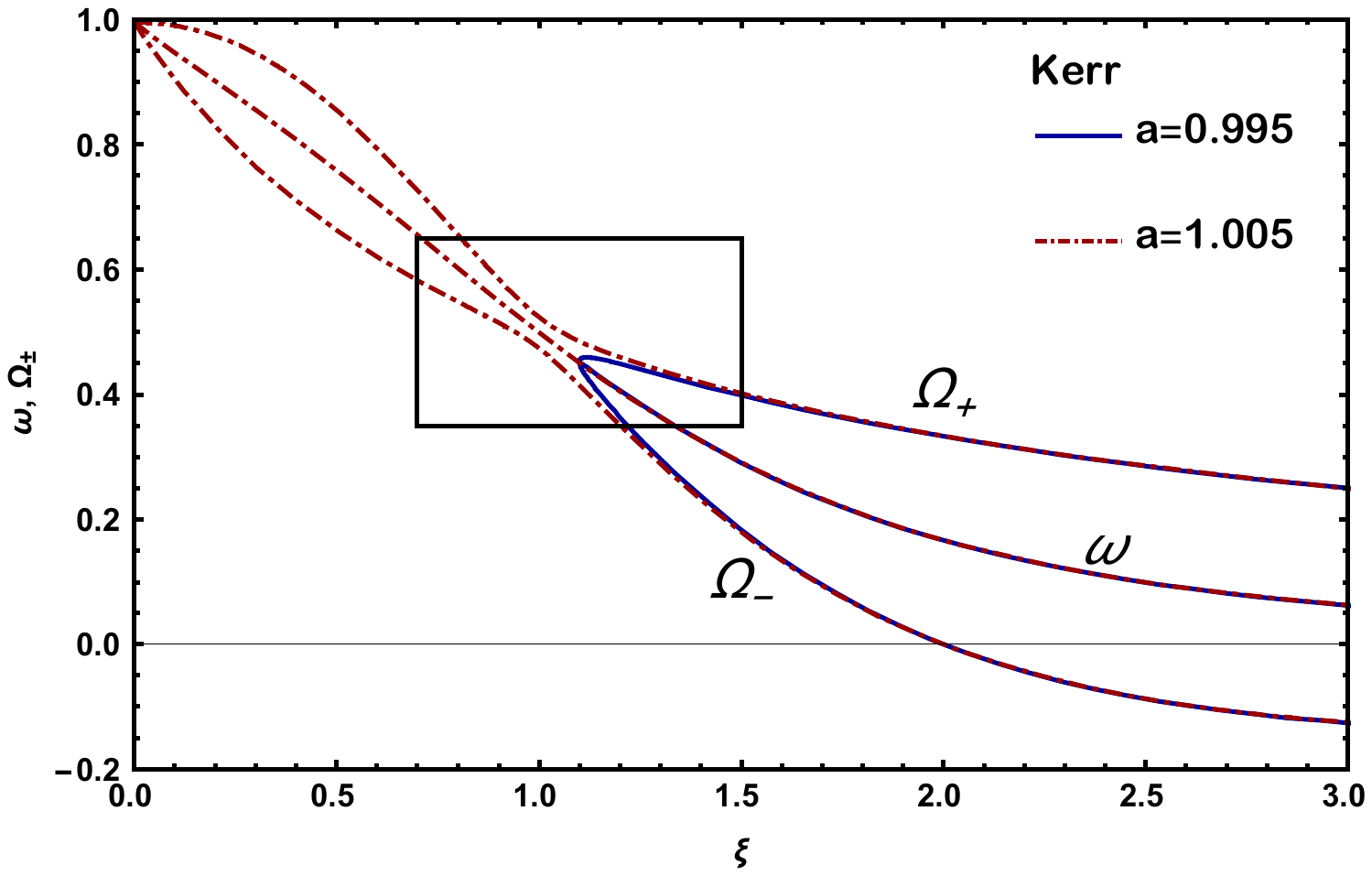}\hfill %
\includegraphics[width=0.49\textwidth,viewport=90 487 519 758]{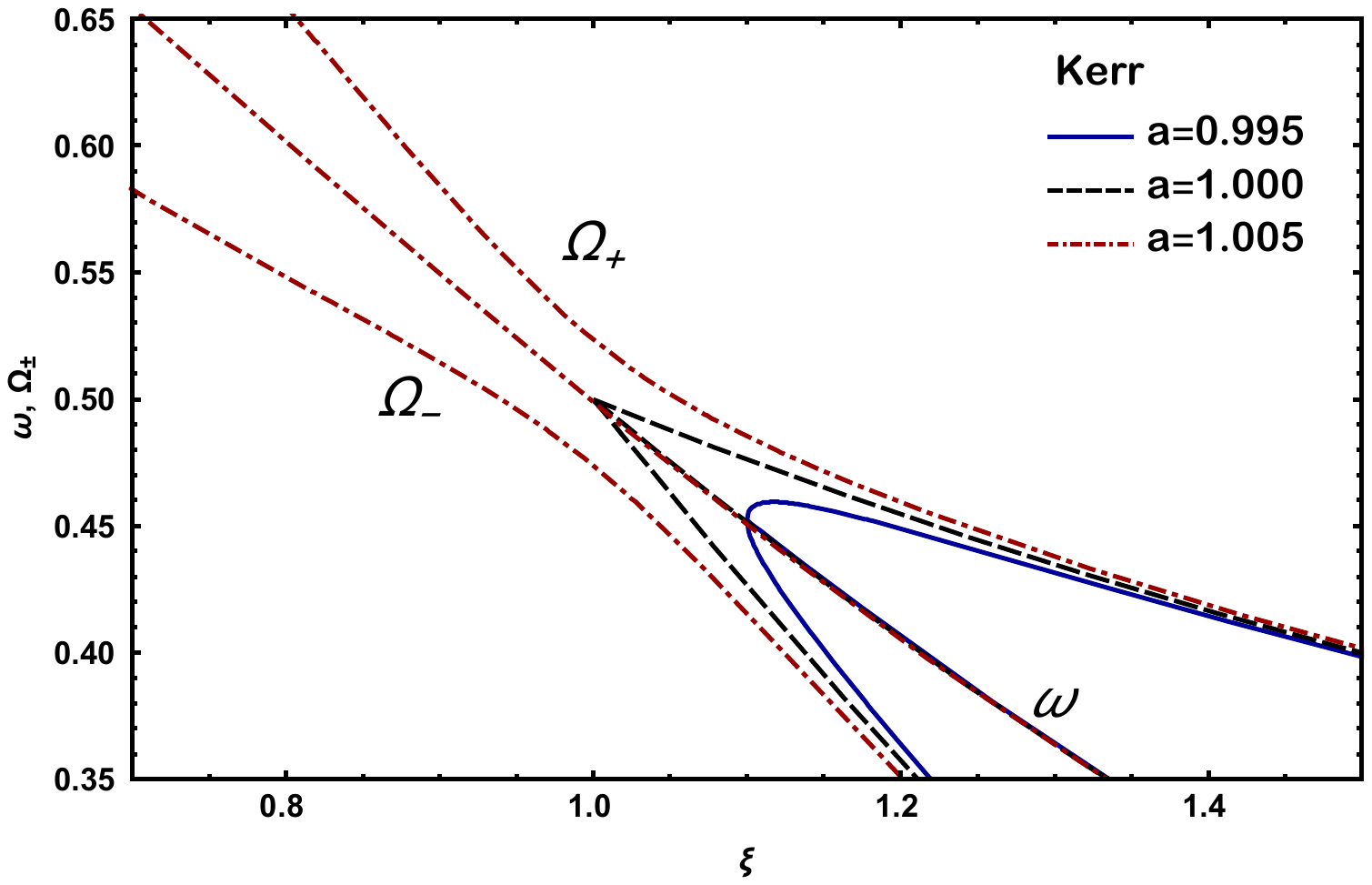}\\[1em]
\includegraphics[width=0.49\textwidth,viewport=90 487 519 758]{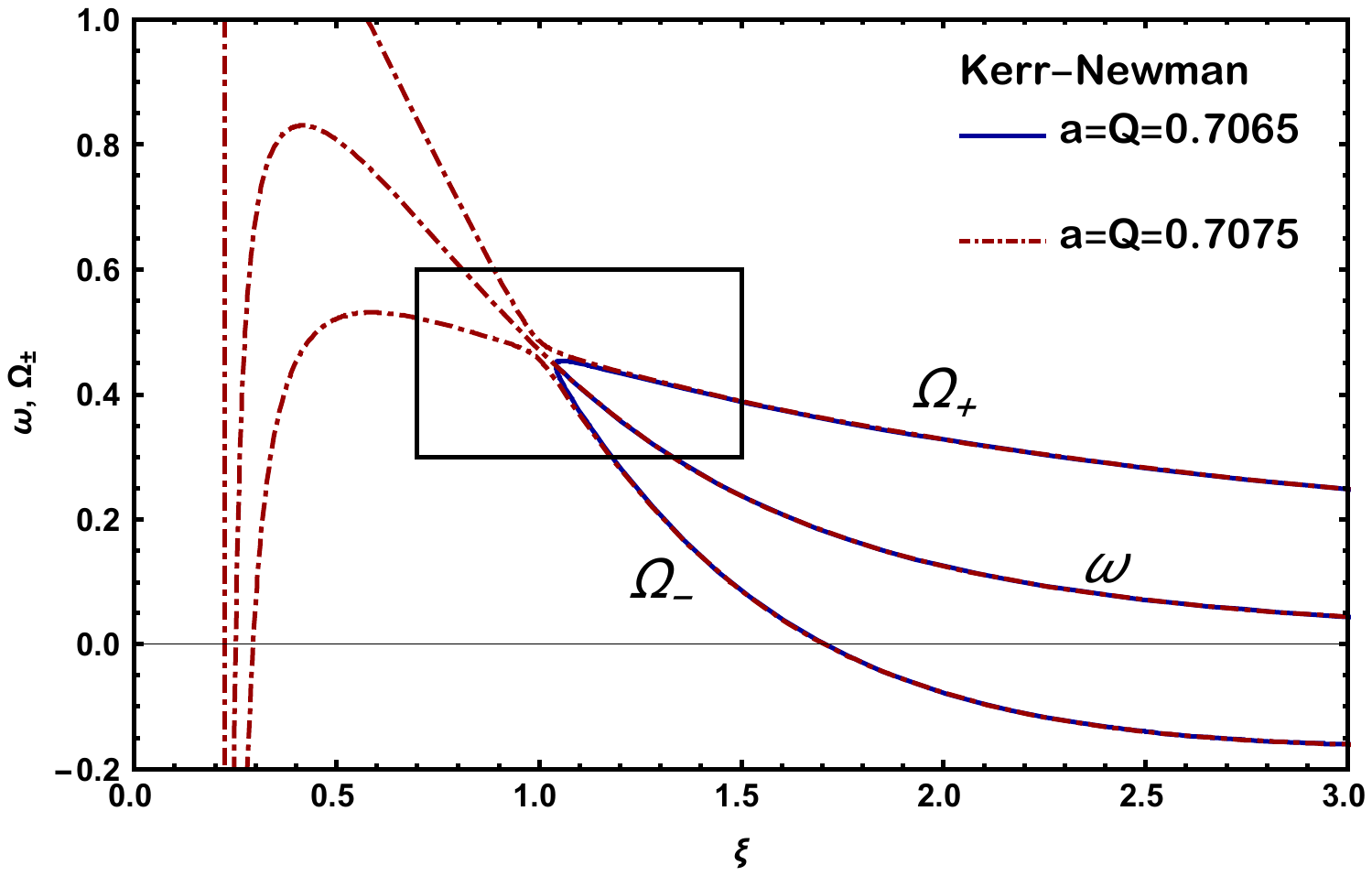}\hfill %
\includegraphics[width=0.49\textwidth,viewport=90 487 519 758]{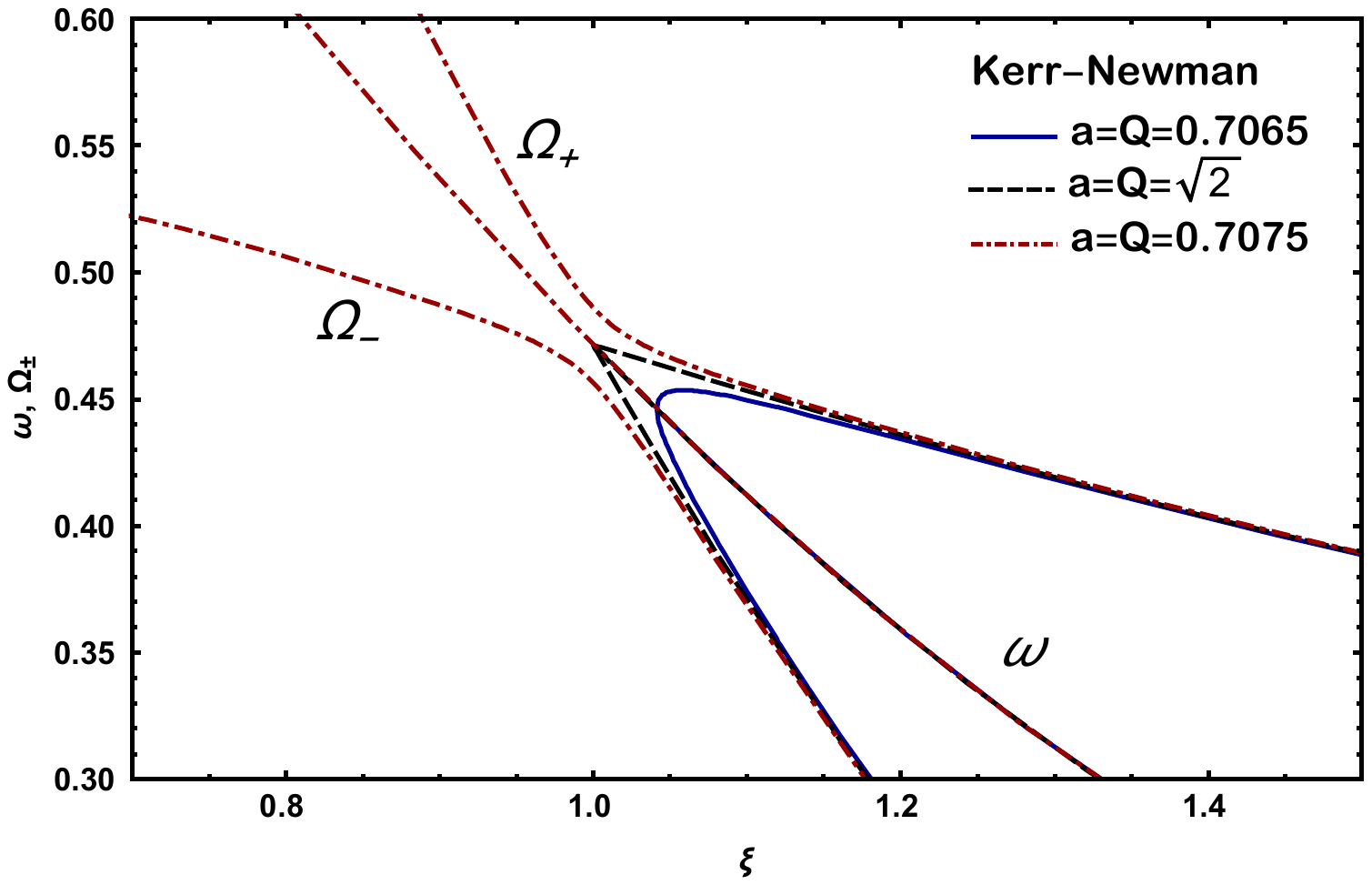}
\caption{\label{Figs} The whale diagram, which shows the three natural angular velocities  for the Kerr and Kerr-Newman metrics. The two figures above are for the Kerr metric, the two below for Kerr-Newman. All graphs are for the natural angular velocities $\Omega_- , \omega, \Omega_+$ as functions of dimensionless radial coordinate $\xi=\rho/\mu$. The graphs on the right show the most interesting regions zoomed in. Solid blue curves show the omegas for near-extremal black holes, dashed black ones correspond to extremal cases (only displayed in the zoomed in versions), dash-dotted red curves are for slightly overspinned naked singularity solutions.  Parameter values are given in the legend. The curves for black holes form a figure that resembles that of a whale, so we call these diagrams ``whale diagrams''. We see that graphs for Kerr and Kerr-Newman differ in the strong field region, but their behavior in the outer region and near the horizon, which determines whether high-energy particles escape to infinity or not, is the same. For generic black hole/singularity spacetimes the graphs can be expected likewise to be modified quantitatively, but not qualitatively.}
\end{figure}

Figure \ref{Figs} shows the natural angular velocities for the near-extremal (solid blue), extremal (dashed black), and slightly overspinned naked singularity (dash-dotted red) Kerr metric and the same graphs for the Kerr-Newman metric (see e.g. \cite{Bicak}) with $a=Q$, as functions of dimensionless radial coordinate $\xi=\rho/\mu$. Here $\rho$ is the Boyer-Lindquist coordinate, which differs from $r$ used in this work by a smooth bounded factor separated from zero; $\mu$ is the black hole's mass. 

On the left we have two sets of curves, for the near-extremal black hole with the dimensionless rotational parameter $a=0.995$, and for the near-extremal naked singularity spacetime with $a=1.005$. On the right the same two sets of curves are shown, plus that for the extremal black hole with $a=1$, in greater detail close to the region $\xi=1$ (the three curves for $\omega$ are overlapping so not all colors are visible). The parameters used for Kerr-Newman are shown on the legend.

We can see that the graphs of the three functions in the Kerr black hole spacetime (solid blue curves on top left graph) form a figure resembling that of a fish, or a whale, which is swimming towards the strong field region, its tail stretching towards the asymptotically flat infinity; its back is composed of $\Omega_+ (r)$, the bottom of $\Omega_- (r)$, middle line of $\omega (r)$. At the ergosurface $\xi=2$ the bottom line crosses the zero level, and the forehead, where all three curves converge, touches the horizon. For the extremal Kerr black hole the forehead of the whale changes into a sharp horn; in the naked singularity spacetime the horn turns into a trunk or tentacle, extending towards the circular singularity $\xi=0$ (this last part is sensitive to the specific metric). Referring to the black hole solution, we call all such graphs the ``whale diagrams''.

The ``trunk'' in the strong field region is different for Kerr-Newman, but the behavior in the small $N$ region and in the outer region, which determines whether high-energy particles escape to infinity or not, is the same. For generic black hole/singularity spacetimes the graphs can be expected likewise to be modified quantitatively, but not qualitatively.

The angular velocity is nice, but it is not an integral of motion. Let us
recast the same inequality in terms of the inverse impact parameter $\beta$,
which can be recovered from (see $u^\mu$ in (\ref{Part4velocity})) 
\begin{equation}
\Omega \equiv \frac{u^\phi}{u^t} =\omega +\frac{L}{X}\frac{N^2}{g_\phi}
=\omega +\frac{\delta\Omega^2}{\beta-\omega},
\end{equation}
and therefore 
\begin{equation}
(\Omega -\omega )(\beta -\omega)=\delta\Omega^2 .
\end{equation}
From this equation we see that, while $\Omega$ lies \emph{inside the whale},
the inverse impact parameter $\beta$ lies \emph{outside the whale}: 
\begin{equation}
(\Omega-\omega) \in \delta\Omega\cdot [-1,1] \quad\Leftrightarrow\quad
(\beta-\omega) \notin \delta \Omega\cdot [-1,1].
\end{equation}
At the same time $\beta$ is an integral of motion, so $\beta$ remains
constant along a geodesic and the whale figure plays the role of an effective potential
graph, which determines the regions where particles with given integrals of
motion can or cannot exist.

The authors of \cite{ph} in this context use $b=\beta^{-1}$ (see Fig. 2 on
p. 10). We think, however, that $\beta$ is more convenient, as this way all
the necessary data is contained in the form of a connected region (the
``whale''), which is only unbounded in the direction of asymptotic infinity.
It should also be noted, that the mentioned Fig.~2 of \cite{ph} can be misleading: the curves $b_+$ and $b_-$ for the naked singularity spacetime seem to intersect there, even though they shouldn't, which is seen clearly on the figure in color. The curves $b_+$ and $b_-$ only intersect at the singularity, where $r=0$, and at the horizon, where $N^2 =0$ (see e.g. Eq. (26) there), if that exists. So for a naked singularity spacetime, in which there is no horizon, there should only be one intersection -- at the
singularity.

Now let us recast $\beta_3$, obtained above, in terms of these parameters: 
\begin{align}
\beta_3 &=\omega +\frac{N}{\sqrt{g_\phi}} \;\frac{\mathcal{E}+P\cos\theta}{%
M\sin\theta}+O(N^2) \\
&=\Omega_+ +\delta\Omega \Big(\frac{\mathcal{E}+P\cos\theta}{M\sin\theta}-1%
\Big) +O(N^2); \\
&=\Omega_+ + \delta\Omega \;\frac{1+\sin (\Phi-\theta)}{\cos\Phi \;\sin\theta%
} +O(N^2),
\end{align}
where $\delta \Omega \sim N$ and 
\begin{equation}
\sin\Phi =\frac{P}{\mathcal{E}}=\tanh \chi =v, \qquad \cos\Phi=\frac{M}{%
\mathcal{E}}=\cosh^{-1}\chi =\gamma^{-1}.
\end{equation}
We see that $(\beta_3 -\Omega_+)$ is positive for particles with $\sin\theta>0$, as it should be, and of the first order in $N$: the particles with diverging energies and momenta are created just above the whale's back ($\Omega_+$) in the region where $N^2$ is small. The qualitative result should have been evident without any
specific calculations.

If we are dealing with a black hole, this means that the highly energetic
particles are created right above the whale's forehead, which is below the
top of its head, and due to the hump -- the maximum of $\Omega_+$ -- they
cannot escape to infinity. Only in the near-extremal case, when the hump is
very low, we can achieve large Killing energies of the escaping particles.
If we are dealing with naked singularity, however, the situation is
different: as $N^2$ does not turn to zero anywhere, the $\Omega_\pm$ curves
converge, but do not intersect. The forehead grows a trunk extending to the
singularity, pointing upward and to the left, with its upper boundary $%
\Omega_+$ monotonically falling away from the region where $N^2$ has
minimum, so the hump is absent and particles a) reflect from the trunk in
the strong field region on the left and b) escape unhindered to infinity on
the right.

\subsection{The bright spot}

The escaping particles all have the value of $\beta_3$ in the zeroth order
by $N$ equal to $\Omega_+ \approx \omega$ at the point of collision, so
their impact parameters tend to 
\begin{equation*}
b_3 \approx \frac{1}{\omega}\Big|_{collision}.
\end{equation*}
Thus the ray of high-energy particles is seen at infinity as emerging from a
``bright spot'' at the equatorial plane, with this impact parameter. As the
particles have large and positive $E_3$ and $L_3$, they are co-rotating with
the black hole or singularity.

For the Kerr near-extremal spacetime $\omega$ at the horizon (or where $N$
is small) is $1/(2M)$, so the impact parameter is $2M$. This is the Eq.
(118) of \cite{ph}.

In the ZAMO frame the tetrad components of a particle's 4-velocity are given
by Eq. (\ref{UTetrad}). Therefore the angle $\theta _{z}$ that the high
energy particles' velocity make with the radial direction in this frame at
the point of collision is given by 
\begin{equation}
\tan \theta _{z}=\frac{u_{3}^{(\varphi )}}{u_{3}^{(r)}}=\frac{L_{3}/\sqrt{%
g_{\varphi }}}{Z_{3}/N}.
\end{equation}%
Using Eq. (\ref{Z}) for $Z_{3}$ with $\epsilon ^{2}=0$ there (for a massless
fragment), we then obtain 
\begin{equation}
	\sin \theta _{z}=\frac{NL_{3}}{X_{3}\sqrt{g_{\phi }}}>0,
	\qquad Z_{3}\cos\theta _{z}>0.
\end{equation}
Using Eqs. (\ref{L3}) for $L_{3}$, (\ref{X3appr}) for $X_{3}$, and (\ref{Mlimit}) for $M$, we obtain 
\begin{equation}
	\sin \theta _{z}\approx \frac{NM}{X_{3}}
		\approx \frac{2\sqrt{\mathrm{X}_{0}^{2}-\mathrm{Z}_{0}^{2}}}
			{\mathrm{X}_{0}+|\mathrm{Z}_{0}|\cos \theta }.
\end{equation}

\subsection{Escape fraction}
We have seen, that the products of the collision at small $N$ for almost all angles $\theta$ have respectively large energies and angular momenta, with their ratio $\beta$, the inverse impact parameter, being close to $\omega$ at the point of collision. From (\ref{E3}) we see that only particles with $(\theta-\theta_E) \sim N$ have $E_3$ and $L_3$ bounded. As $\theta_E$ is also of the order of $N$ (\ref{SmallAngles}), this means that only particles created in a narrow cone $\theta \sim N$ around radial direction do not have large Killing energies and angular momenta. Those that are not created in this narrow cones have diverging $E_3$ and $L_3$, either positive or negative, with almost the same value of impact parameter $b_3$.

Let us assume for simplicity that the distribution of the particles in the center of mass frame is independent of the small parameter $N$ and is isotropic. Then almost half of the created particles, with $\sin\theta>0$, are created with large positive $E_3$ and $L_3$. According to the previous section, they are situated \emph{just above} the forehead/horn/trunk on the whale diagram. In the black hole case with the forehead, they cannot escape to infinity due to the potential barrier, but in an overspinned spacetime with the trunk they can and do escape to infinity (we do not discuss here the case of extremal black holes).

The other half of the particles, with $\sin\theta<0$, has large negative $E_3$ and $L_3$. On the whale diagram they are situated \emph{just below} the trunk and can only travel in the direction of smaller $r$.

This is what happens when the distribution of created particles in the center of mass frame is isotropic, as assumed in \cite{ph}. This may not be very realistic assumption. The two colliding particles in the center of mass frame have large energies and momenta along the radial direction (\ref{ur}). Therefore realistic distributions can be highly anisotropic, with the distinguished radial direction, depending on the specific particles colliding. This issue could be clarified if the detailed mechanism of interaction between colliding  particles is taken into account, but this problem is far beyond the scope of our paper.
%If most of created particles are created in a narrow cone close to the radial direction, then more of them will have $\theta\sim N$ and bounded Killing energies. Suppose $\theta_a$ is the angle, such that half of the particles created in the collision are created within the cone $\theta<\theta_a$. If $\theta_a \sim \gamma^{-1}\sim N$, then the escape fraction will still be of the order of unity. In case $\theta_a$ is of the higher order by $\gamma$, the escape fraction will be suppressed by a corresponding large factor, but in case the angle $\theta_a (N)$ is separated from zero or just not that small, the escape fraction will not be significantly affected.

The value of the escaping fraction was obtained in \cite{ph} through detailed calculations of allowed angles depending on the sign of $\mathrm{Z}_0$. Although these details can be of interest on their own, we see that, inasmuch as we are interested in the escape fraction in the large $M$ limit, they are not necessary.

\section{Conclusion}

We derived general formulas which describe the process of collision of two
massive particles producing two massless ones. It was shown that the relation between the energy in the center of mass frame $E_{c.m.}$ and the Killing energy of a fragment $E$ follows immediately from the Wald inequalities; that divergence of $E_{c.m.}$ is a necessary, but not sufficient, condition for the divergence of $E$. This, in particular, enabled us to understand the previous results for the particle collisions near the wormhole throat, when the strong inequality $E\gg E_{c.m.}$ can take place. The connection between the details of the collision process in the stationary and the ZAMO (LNRF) frames was analyzed extensively, and the geometry of the escape cone was described in the general setting. The existence of a bright spot follows immediately from those general formulas. We also found the escape fraction of the fragments after collision. In the particular case of the Kerr metric our results agree with those of \cite{ph}.

\section*{Acknowledgments}
The work of O. Z. was funded by the subsidy allocated to Kazan Federal
University for the state assignment in the sphere of scientific activities.

%\appendix
\section*{Appendix A. Derivation of formulas for $L_3$ and $E_3$}
Here we present in more detail the calculations for the integrals of motion $E$ and $L$ of one of the fragments (\ref{L3}--\ref{CosT'}) in terms of the collision parameters in the center of mass frame, $M$ and $\theta$, and the tetrad components of the frame's momentum in the LNRF frame (\ref{A})--(\ref{B}), as presented in section \ref{sec:LE3}.

We start from Eq. (\ref{L30}) for $L_3$. It is a linear combination of $\sin\theta$ and $\cos\theta$, which can be simplified step by step by using the expressions for $\chi$ (\ref{chi}) and $\psi$ (\ref{psi}) as follows:
\begin{align}
\frac{L_3}{\sqrt{g_\phi}}&=\frac{M}{2} 
	\big[\sinh\chi \sin\psi +\cosh\chi \sin\psi \cos\theta +\cos\psi \sin\theta \big] \\
&=\frac{\mathrm{L}_0 /2}{\sqrt{g_\phi}}
	+\frac{1}{2} \big[M\cos\psi \sin\theta +\mathcal{E}\sin\psi \cos\theta \big] \\
&=\frac{\mathrm{L}_0 /2}{\sqrt{g_\phi}}
	+\frac{1}{2} \Big[M\frac{P_r}{P}\sin\theta 
		+\mathcal{E}\frac{P_\phi}{P}\cos\theta \Big] \\
&=\frac{\mathrm{L}_0 /2}{\sqrt{g_\phi}}
	+\frac{1}{2}\sqrt{M^2 +P_\phi^2}\, \sin (\theta-\theta_L),
\end{align}
where 
\begin{equation}
\cos\theta_L =\frac{P_r}{P}\frac{M}{\sqrt{M^2 +P_\phi^2}},\qquad
\sin\theta_L =-\frac{P_\phi}{P}\frac{\mathcal{E}}{\sqrt{M^2 +P_\phi^2}}.
\end{equation}
This can be more elegantly written in terms of 
\begin{align}
P_{\phi 3}\equiv \frac{L_3}{\sqrt{g_\phi}}
	=& \tfrac12 P_\phi +\tfrac12 \sqrt{P_\phi^2 +M^2}\;\sin (\theta-\theta_L), \\
&=\frac{\mathrm{L}_0 /2}{\sqrt{g_\phi}}
	+ \frac{\sqrt{\mathrm{X}_0^2 -\mathrm{Z}_0^2}}{2N}\;\sin(\theta-\theta_L), \\
& \sin\theta_L =-\frac{P_\phi /P}
	{\sqrt{1 -\mathrm{Z}_0^2 /\mathrm{X}_0^2}}, \quad P_r \cos\theta_L >0 ,
\end{align}
which brings us to (\ref{L3})--(\ref{CosTL}).

The expression for energy can then be transformed in the following way: from the definition of $X_3 = E_3 -\omega L_3$, which is given by Eq. (\ref{X3}), we get
\begin{align}
E_3 &= X_3 +\omega L_3 
	=\frac{\mathrm{X}_0}{2}+\omega \sqrt{g_\phi}\; P_{\phi3}
		+\frac{PN}{2}\cos\big(\theta_L +(\theta-\theta_L)\big) \\
&=\frac{\mathrm{E}_0}{2}+\frac{N/2}{\sqrt{M^2 + P_\phi^2}} 
	\Big\{MP_r \cos(\theta-\theta_L) +\notag\\
		&\qquad\qquad\qquad\qquad+\Big(\mathcal{E}P_\phi +\omega\sqrt{\frac{g_\phi}{N^2}}\; 
			\big[M^2 +P_\phi^2 \big]\Big)\sin(\theta-\theta_L)\Big\} \notag \\
&=\frac{\mathrm{E}_0}{2}
	+\frac{N}{2}\sqrt{\frac{M^2 P_r^2 
		+\Big(\mathcal{E}P_\phi +\omega\sqrt{\frac{g_\phi}{N^2}}
			\;\big[M^2 +P_\phi^2 \big]\Big)^2 }{M^2+P_\phi^2}}\;
				\sin(\theta-\theta_L-\theta^{\prime}), \notag\\
&=\frac{\mathrm{E}_0}{2}+\frac{W}{2}\sin(\theta -\theta_L -\theta^{\prime}),
\end{align}
where the amplitude $W$ is further simplified
\begin{align}
W^2&=N^2 \frac{M^2 P_r^2 +\Big(\mathcal{E}P_\phi +\omega\sqrt{\frac{g_\phi}{%
N^2}}\;\big[M^2 +P_\phi^2 \big]\Big)^2 }{M^2 +P_\phi^2} \\
&=\big(\omega \sqrt{g_\phi}\,\mathcal{E}+NP_\phi\big)^2 -g_t P_r^2 \\
&=\mathrm{E}_0^2 +g_t M^2 ; 
\end{align}
and the new angle $\theta'$ is given by
\begin{align}
\sin\theta^{\prime}&
	=-\frac{\sqrt{g_\phi}\; \mathrm{Z}_0 M}{W\Lambda};  
			\label{SinT'a}\\
\cos\theta^{\prime}&
	=\frac{\mathrm{L}_0 \mathrm{E}_0 +\omega g_\phi M^2}{W\Lambda}.
			\label{CosT'a}
\end{align}

Using the expressions for $\sin\theta_L$, $\cos\theta_L$, $\sin\theta^{\prime}$, $\cos\theta^{\prime}$ from (\ref{SinTL},\ref{CosTL},\ref{SinT'},\ref{CosT'}), the definitions of $\mathcal{E},P,P_r, P_\phi, M$ from Eqs. (\ref{A}--\ref{M}), and some more algebra, we can obtain $\theta_E \equiv \theta_L +\theta^{\prime}$: 
\begin{align}
	\sin\theta_E \equiv \sin (\theta_L +\theta^{\prime})
		&=\frac{NM^2 -\mathrm{E}_0 \mathcal{E}}{PW}; \\
	\cos\theta_E \equiv \cos (\theta_L +\theta^{\prime})
		&=\frac{P_r \cdot \omega \sqrt{g_\phi}\;M}{PW}.
\end{align}
Gathering together the expression for $E_3$, we arrive to (\ref{E3})--(\ref{CosTE}). 

\section*{Appendix B. Expansions at small $N$}
In this section we provide more detailed derivation of the expansions for various quantities of interest, used in subsection \ref{sec:limit}, for $N\ll 1$.

\setcounter{paragraph}{0}
\paragraph{Amplitudes.}

For the angular momentum amplitude we have 
\begin{align}
	\Lambda&=\sqrt{g_\phi}\; M 
		\Big(1+\frac{\mathrm{L}_0^2}{g_\phi}\,\frac{1}{2M^2}+O(N^4)\Big); 
\end{align}
and for the energy
\begin{align}
	W&=\sqrt{g_t}\; M 
		\Big(1+\tfrac12 \frac{\mathrm{E}_0^2}{g_t M^2}+O(N^4)\Big) \\
		&=\omega \sqrt{g_\phi}\; M 
		\Big(1-\tfrac12 \frac{N^2}{g_\phi \omega^2} 
			+\tfrac12 \frac{\mathrm{E}_0^2}{g_\phi \omega^2 M^2}+O(N^4)\Big) \\
	&=\omega \sqrt{g_\phi}\; M 
		\Big(1 +\frac{\mathrm{E}_0^2 +\mathrm{Z}_0^2 -\mathrm{X}_0^2}
			{g_\phi \omega^2}\,\frac{1}{2M^2}+O(N^4)\Big).
\end{align}

\paragraph{Angles.}

The three angles are related by $\theta^{\prime}=\theta_E -\theta_L$, so of
6 functions for their sines and cosines it is sufficient to derive only two
asymptotes (and bear in mind all the signs). For $\theta_L$ we get 
\begin{align}
\sin\theta_L &=-\frac{\mathrm{X}_0}{M} 
	\frac{P_\phi}{\sqrt{\mathrm{Z}_0^2 +P_\phi^2 N^2}} 
		\Big(1-\frac{\mathrm{L}_0^2}{g_\phi}\,\frac{1}{2M^2}+O(N^4) \Big) \\
	&=-\gamma_0 \sin\psi 
		\Big(1-\frac{\mathrm{L}_0^2}{g_\phi}\,\frac{1}{2M^2}+O(N^4) \Big) \sim N ; \\
	\cos\theta_L &\approx \text{sign}(\mathrm{Z}_0)\equiv \sigma.
\end{align}
Likewise for $\theta^{\prime}$ 
\begin{align}
\sin\theta^{\prime}&=-\frac{\mathrm{Z}_0}{\omega \sqrt{g_\phi}\;M} 
	\Big(1-\tfrac12 \frac{\mathrm{L}_0^2}{g_\phi M^2} 
		-\frac12 \frac{\mathrm{E}_0^2 +\mathrm{Z}_0^2 -\mathrm{X}_0^2}
			{g_\phi \omega^2 M^2}+O(N^4)\Big) \\
&= -\frac{\mathrm{Z}_0}{\omega \sqrt{g_\phi}\;M}
	\Big(1-\frac{\mathrm{Z}_0^2 +2\omega \mathrm{E}_0 \mathrm{L}_0}
		{g_\phi \omega^2}\, \frac{1}{2M^2}+O(N^4) \Big) \\
&= -\frac{\mathrm{Z}_0}{M}\;\frac{1}{\omega \sqrt{g_\phi}} 
	\Big(1-\frac{\mathrm{Z}_0^2 +2\omega \mathrm{E}_0 \mathrm{L}_0}
		{g_\phi \omega^2}\, \frac{1}{2M^2}+O(N^4) \Big) \\
&= -\frac{N}{\omega\sqrt{g_\phi}} \;v_0 \gamma_0 \cos\psi 
	\Big(1-\frac{\mathrm{Z}_0^2 +2\omega \mathrm{E}_0 \mathrm{L}_0}
		{g_\phi \omega^2}\, \frac{1}{2M^2}+O(N^4) \Big)\sim N; \\
\cos\theta^{\prime}&\approx +1,\qquad \theta_E \approx \theta_L
\end{align}

Thus all angles $\theta_L. \theta_E, \theta^{\prime}$ are small and of the
order of $N$. There is a narrow cone with the angle $O(N)$ around the radial
direction $\theta=0$, in which $E_3$ and $L_3$ are finite despite diverging $%
M$, while for all other angles both $E_3$ and $L_3$ diverge as $M\sim 1/N$.

In the generic situation $\mathrm{Z}_0 \neq 0$, so we have $\theta_L \sim \theta^{\prime}\sim \theta_E \sim N$ (\ref{SmallAngles}).
%\begin{equation*}
%\theta_L \sim \theta^{\prime}\sim \theta_E \sim N .
%\end{equation*}

\paragraph{Impact parameter.}

When both $\Lambda$ and $W$ diverge, and $\theta$ is not in the narrow cone $%
\sim O(N)$, the impact parameter tends to 
\begin{equation}
b_3 =\frac{L_3}{E_3}
	=\frac{\mathrm{L}_0 +\Lambda \sin (\theta-\theta_L)}
		{\mathrm{E}_0 +W\sin(\theta-\theta_E)} 
			\approx \frac{\Lambda}{W}\approx \frac{1}{\omega}.
\end{equation}
Let us find the asymptotic behavior in the next order of magnitude. 
\begin{equation}
b_3 =\frac{\Lambda \sin (\theta-\theta_L)}{W\sin(\theta-\theta_E)} 
	\Big(1+\frac{\mathrm{L}_0}{\Lambda}\sin^{-1}(\theta-\theta_L)
		-\frac{\mathrm{E}_0}{W}\sin^{-1}(\theta-\theta_E) 
			+O\Big(\frac{\mathrm{E}_0}{W}\Big)^2\Big).
\end{equation}
With the help of
\begin{align}
\sin (\theta-\theta_L)&=\sin (\theta-\theta_E)
\cos\theta^{\prime}+\cos(\theta-\theta_E)\sin \theta^{\prime} \\
&=\sin (\theta-\theta_E) \big[1+\sin\theta^{\prime}\cot (\theta-\theta_E) %
\big]+O(N^2), \\
\frac{1}{\sin (\theta-\theta_L)}&=\frac{1}{\sin (\theta-\theta_E)} \big[%
1-\sin\theta^{\prime}\cot (\theta-\theta_E) \big]+O(N^2),
\end{align}
this can be transformed to 
\begin{align}
b_3 &=\frac{\Lambda}{W}
	\Big[1 +\sin \theta^{\prime}\cot (\theta-\theta_E)+O(N^2)\Big] \\
&\qquad\times
		\Big(1+\frac{\mathrm{L}_0}{\Lambda}\sin^{-1}(\theta-\theta_L) 
		-\frac{\mathrm{E}_0}{W}\sin^{-1}(\theta-\theta_E) +O(N^2)\Big) \\
&=\frac{1}{\omega}\big(1+O(N^2)\big) 
	\Big[1-\frac{\mathrm{Z}_0}{M\omega \sqrt{g_\phi}}
		\cot (\theta-\theta_E)+O(N^2)\Big]	\times \\
&\qquad \times \Big(1
	+\frac{\mathrm{L}_0}{\sqrt{g_\phi}\;M}\sin^{-1}(\theta-\theta_L)
	-\frac{\mathrm{E}_0}{\omega\sqrt{g_\phi} M}
		\sin^{-1}(\theta-\theta_E) +O(N^2)\Big) \\
&=\frac{1}{\omega} 
	\Big[1-\frac{\mathrm{Z}_0}{M\omega \sqrt{g_\phi}}
		\cot (\theta-\theta_E)+O(N^2)\Big]\times \\
&\qquad \times \Big(1
	+\frac{\sin^{-1}(\theta-\theta_E)}{\omega\sqrt{g_\phi}\;M} 
		\big[-\mathrm{E}_0 +\omega \mathrm{L}_0 (1-O(N))\big] +O(N^2)\Big) \\
&=\frac{1}{\omega} 
	\Big[1-\frac{\mathrm{Z}_0}{M\omega \sqrt{g_\phi}}\cot(\theta-\theta_E)
		+O(N^2)\Big]\\
&\qquad\times 
	\Big(1-\frac{\mathrm{X}_0}{\omega\sqrt{g_\phi}\;M}
		\sin^{-1}(\theta-\theta_E) +O(N^2)\Big) \\
&=\frac{1}{\omega}\Big[1-\frac{N}{\omega \sqrt{g_\phi}} 
	\frac{\mathcal{E}+P_r \cos(\theta-\theta_E)}
		{M\sin (\theta-\theta_E)}+O(N^2)\Big] <\frac{1}{\omega}.
\end{align}
We should remember, that all the metric coefficients here are taken at the
point of collision. 

Now suppose that 
\begin{equation*}
\omega =\omega_0 -\omega_1 N+O(N^2),
\end{equation*}
where $\omega_0$ is the value of $\omega$ at the point where $N^2 =0$. Then 
\begin{equation}
b_3 =\frac{1}{\omega_0}\Big[1+N \Big(\frac{\omega_1}{\omega_0} -\frac{%
\mathcal{E}+P_r \cos(\theta-\theta_E)}{\sqrt{g_t}\;M\sin (\theta-\theta_E)}%
\Big)+O(N^2)\Big].
\end{equation}

\end{document}